\documentclass{PoS}
\usepackage{amsmath,amssymb}

\newcommand{\tmtextbf}[1]{{\bfseries{#1}}}

\newcommand{\tmtexttt}[1]{{\ttfamily{#1}}}

\newcommand{\bea}{\begin{eqnarray}}
\newcommand{\eea}{\end{eqnarray}}
\newcommand{\eq}[1]{Eq.~(\ref{#1})}
\newcommand{\fig}[1]{Fig.~\ref{#1}}

\def\tr{{\rm tr}}
\newcommand{\plaq}{{\rm Plaq}}
\newcommand{\pq}[1]{\plaq_{#1}}

\def\aqq{\alpha_{\rm qq}}

\def\Lmin{L^{\rm min}}
\def\Ls{L_{\rm s}}

\def\L5min{\Lmin_5}

\newcommand{\ov}{\overline v}
\newcommand{\oh}{\overline h}

\title{Dimensional reduction from five-dimensional gauge theories}

\ShortTitle{Dimensional reduction from five-dimensional gauge theories}

\author{\speaker{Francesco Knechtli} and Antonio Rago\footnote{
        Present address: School of Computing and Mathematics, University of
        Plymouth, Plymouth PL4 8AA, UK}\\
        Department of Physics, Bergische Universit\"at Wuppertal\\
        Gaussstr. 20 \\ 
        D-42119 Wuppertal, Germany \\
\\
        E-mail: \email{knechtli@physik.uni-wuppertal.de}}

\abstract{
We study the phase diagram of five-dimensional SU(2) gauge theories on
anisotropic lattices with periodic boundary conditions. We locate a line
of first order bulk phase transitions and second order phase transitions
related to breaking of the center along one direction. A reduction to four
dimensions occurs through compactification of one dimension but not only.
By choosing a lattice spacing in the extra dimension larger than in the
other dimensions, we find hints that the hyperplanes orthogonal to the
extra dimension decouple from each other. Our analysis is based on
measurements of Polyakov loops, the static potential extracted from Wilson
loops and renormalized couplings defined through the static force and its
derivative.
\begin{flushright} WUB/11-20 \end{flushright}
}

\FullConference{XXIX International Symposium on Lattice Field Theory \\
                 July 10-16 2011\\
                 Squaw Valley, Lake Tahoe, California}

\begin{document}
\section{Introduction \label{s_ic}}

Our interest in five-dimensional gauge theories is motivated by the models of
physics beyond the Standard Model called Gauge-Higgs Unification. When
dimensional reduction from five to four dimensions occurs, the
five-dimensional components of the gauge field behave as scalar fields which
could be identified with the Higgs particle. The physical content of the
scalar fields is carried by the Polyakov loops winding along the fifth
dimension. In this contribution we concentrate on the possibility of
dimensional reduction and the mechanisms underlying it. We emphasize that due
to the non-renormalizability (or triviality) of five-dimensional gauge
theories, it is important to perform this study in five dimensions, despite
the computational cost. In fact triviality is related to the existence of a
bulk phase transition in five dimensions \cite{Creutz:1979dw}.

Our setup is a $L_T\times \Ls^3\times L_5$ Euclidean lattice. The SU(2)
Yang--Mills theory is discretized using the
anisotropic Wilson plaquette gauge action \cite{Wilson:1974sk}
\bea
S & = & \frac{\beta}{2} \sum_x \left[ \frac{1}{\gamma} \sum_{\mu<\nu}
  \tr\{1-U(x;\mu,\nu)\} + \gamma \sum_{\mu} \tr\{1-U(x;\mu,5)\} \right] \,,
\eea
where the traces are over oriented plaquettes and the indices $\mu,\nu$ run
over the usual four dimensions. The lattice couplings $\beta$ and $\gamma$ are
are related in the classical limit to the lattice spacings $a_4$ in four
dimensions, $a_5$ in the fifth dimension and to the dimensionful bare gauge
coupling $g_5$ through
\bea
\beta = \frac{4a_4}{g_5^2} \quad & \mbox{and} & 
\quad \gamma = \frac{a_4}{a_5} \,.
\eea
An alternative but equivalent parameter pair is
\bea
\beta_4 = \frac{\beta}{\gamma} \quad  & \mbox{and} & 
\quad \beta_5 = \beta\,\gamma \,.
\eea
We take periodic boundary conditions in all the directions. Here we discuss
the main results and refer to \cite{Knechtli:2011gq} for more details.
In \cite{Yoneyama} we report on recent results using orbifold boundary
conditions.

\section{The mean-field phase diagram \label{s_mf}}
\begin{figure}\centering
  \resizebox{10cm}{!}{\includegraphics[angle=0]{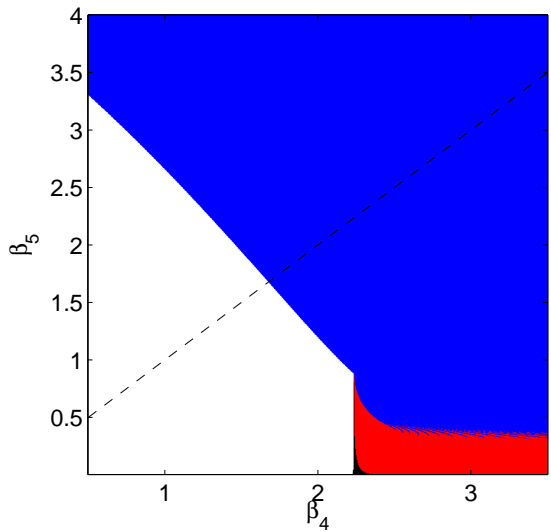}}
  \caption{The phase diagram of the five-dimensional SU(2) gauge theory
    using the anisotropic Wilson plaquette gauge action in the mean-field
    approximation (saddle-point solution).}
  \label{f_meanfield}
\end{figure}

In \cite{Irges:2009bi,Irges:2009qp} the five-dimensional SU(2) gauge theory
has been investigated using the mean-field method, which is an expansion
around a saddle-point. The saddle-point configuration is parametrized by
$\ov{\bf 1}$ for the four-dimensional links and $\ov_5{\bf 1}$ for extra
dimensional links (${\bf 1}$ is the $2\times2$ identity matrix) and is found
by iteratively solving the equations
\bea
\ov = \frac{I_2(\oh)}{I_1(\oh)} \quad & \mbox{and} &
\quad \ov_{5} = \frac{I_2(\oh_{5})}{I_1(\oh_{5})} \,,
\eea
where $I_{1,2}$ are modified Bessel functions and $\oh=6\beta_4\ov^3
+2\beta_5\ov\ov_5^2$, $\oh_5=8\beta_5\ov^2\ov_5$. The result is shown in
\fig{f_meanfield}. There are three phases, one in which $\ov\neq0$,
$\ov_5\neq0$ (blue area, the deconfined phase), one in which $\ov\neq0$,
$\ov_5=0$ (red area, the layered phase) and one in which $\ov=0=\ov_5$ (white
area, the confined phase). The black points mark values where no convergence
to a finite value is attained and the dashed line represents
$\beta_4=\beta_5$. 

In \cite{Irges:2009qp} it was found that
dimensional reduction occurs when $\beta_4>\beta_5$, or $a_5>a_4$. The
transition from the deconfined to the layered phase turns second order and a
continuum limit can be taken. In the limit, the four-dimensional hyperplanes
orthogonal to the fifth dimension decouple and a theory in the
universality class of the four-dimensional Ising model is obtained. The
correlation length is given by the inverse scalar mass. The continuum limit is
taken in finite volume keeping two quantities, the anisotropy $\gamma$ and the
ratio of the vector to the scalar mass, fixed.
In the following we verify if
this scenario is confirmed by Monte Carlo simulations of the full theory.

\section{The phase diagram from Monte Carlo simulations \label{s_mc}}
\subsection{Bulk phase transitions}
\begin{figure}\centering
  \resizebox{12cm}{!}{\includegraphics[angle=-90]{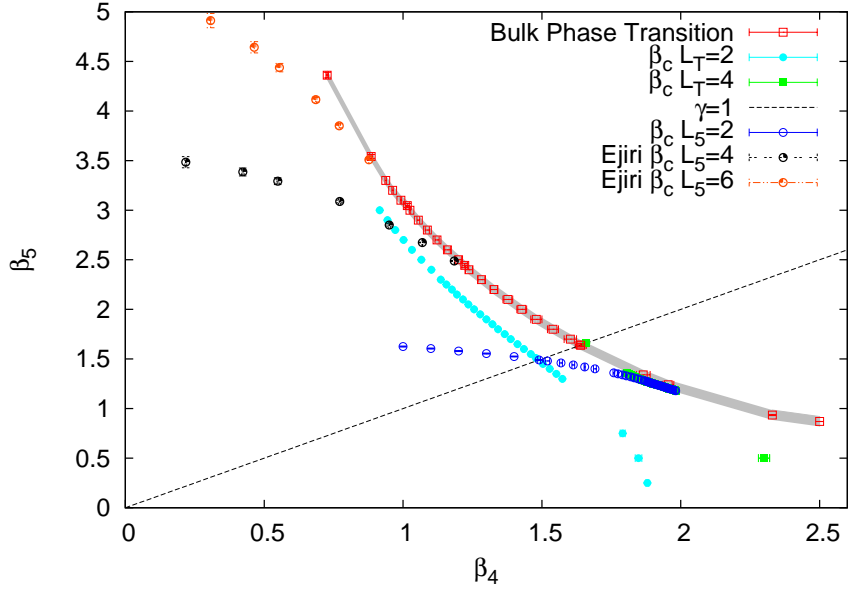}}
  \caption{The phase diagram of the five-dimensional SU(2) gauge theory
    simulated with the anisotropic Wilson plaquette gauge action
    \cite{Knechtli:2011gq}.}
  \label{f_phasediag}
\end{figure}

\begin{figure}\centering
  \resizebox{10cm}{!}{\includegraphics[angle=-90]{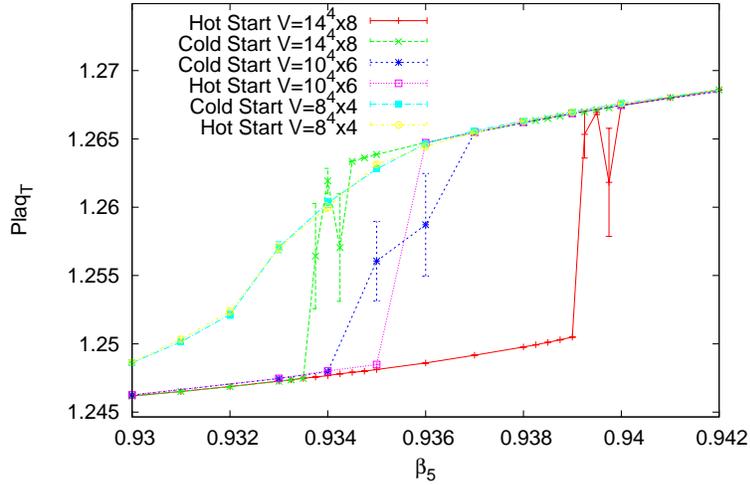}}
  \caption{Bulk phase transition at $\beta_4=2.33$. Lattices with $\Ls\ge14$
    are needed to see the hysteresis in the plaquette (here $\pq{T}$ is the
    four-dimensional plaquette).}
  \label{f_hysteresis}
\end{figure}

\fig{f_phasediag} is the summary of our results concerning the phase diagram
\cite{Knechtli:2011gq}. The points marked by red squares (and connected by a
grey band to guide the eye) are bulk phase transitions and they are signaled
by the behavior of the plaquette.
\fig{f_hysteresis} shows this transition in a scan of $\beta_5$
keeping $\beta_4=2.33$ fixed. The hysteresis effect is seen provided the
four-dimensional volume is large enough, in this case we need $L_T=\Ls\ge14$ 
($L_5$ is approximately $\Ls/2$).  If the volume is smaller the transition
appears like a cross-over, due to compactification of these directions
(we will return to this issue in the next section).
In order to study bulk phase transitions when
$\beta_4>\beta_5$ we need to ensure that four directions of the lattice ($L_T$,
$\Ls$) are large enough, since $a_4<a_5$. This has to be compared with studies
at $\beta_5>\beta_4$, where $L_5$ only has to be made sufficiently large. In
summary, the grey band in \fig{f_phasediag} is a line of first order phase
transitions. From measurements of the static potential we know that this line
separates the confined phase at smaller $\beta_4$ from the deconfined
phase.

\subsection{Phase transitions related to center breaking \label{s_cb}}
\begin{figure}\centering
  \resizebox{7cm}{!}{\includegraphics[angle=-90]{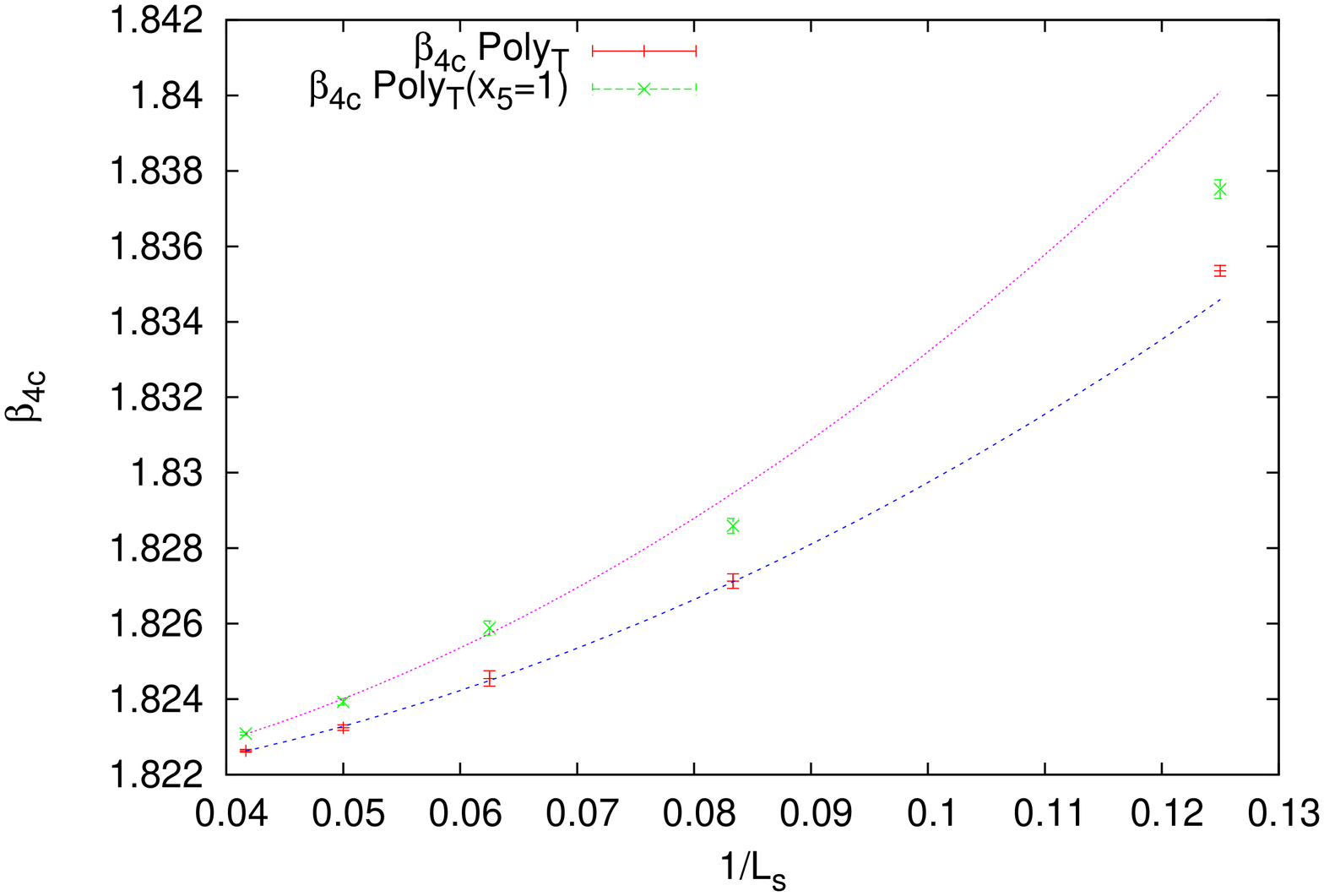}} \ \
  \resizebox{7cm}{!}{\includegraphics[angle=-90]{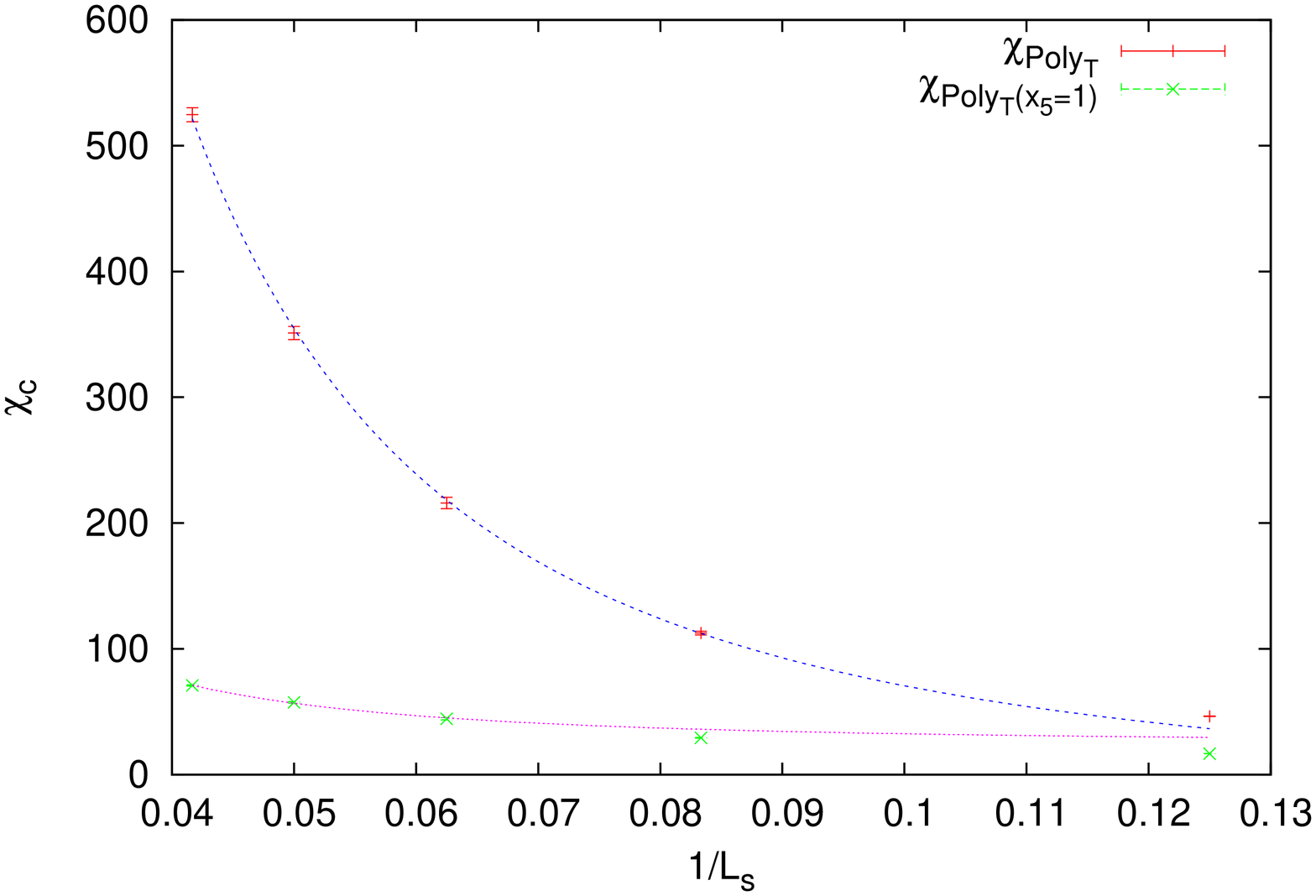}}
  \caption{Finite size scaling analysis for the transition at
    $\beta_5=0.5$, $L_T=2$: the critical coupling $\beta_{4c}$ (left plot) and
    the critical susceptibility $\chi_c$ of the temporal Polyakov loop (right
    plot). The lines are fits to
    the data using the critical exponents of the four-dimensional Ising model.
    Two definitions of the Polyakov loop are used, see \cite{Knechtli:2011gq}.}
  \label{f_universality}
\end{figure}

\fig{f_phasediag} presents phase transition points due to compactification of
one direction, which are signaled by the behavior of the Polyakov loop winding
along the small direction. When compactification occurs, the Polyakov loop
expectation value (its absolute value) becomes non-zero. At $\beta_5>\beta_4$
there are second order phase transitions when $L_5=2,4,6,\ldots$ which have
been studied in \cite{Ejiri:2000fc,deForcrand:2010be}. Our new results mainly
concern the region $\beta_4>\beta_5$, where we compactify
$L_T=2,4,\ldots$. \fig{f_universality} shows a study of the
transition with $L_T=2$ at $\beta_5=0.5$. We do a finite size scaling
analysis varying $\Ls=8\ldots24$ while keeping $L_5=\Ls/2$. We measure the
critical coupling $\beta_{4c}(\Ls)$, at which the susceptibility of the temporal
Polyakov loop has its maximum $\chi_c(\Ls)$. The data for $\beta_{4c}$ and
$\chi_c$ are perfectly compatible with the scaling laws
\bea
|\beta_{4c}(\Ls)-\beta_{4c}(\Ls=\infty)| &\sim& \Ls^{-1/\nu} \quad
   \mbox{(4d Ising: $\nu=1/2$)} \,,
\\
\chi_c(\Ls) &\sim& L^{\gamma/\nu} \quad \mbox{(4d Ising: $\gamma=1$)} \,,
\eea
using the values of the critical exponents $\nu$, $\gamma$ of the
four-dimensional Ising model. We thus confirm \cite{Svetitsky:1982gs}.
\begin{figure}\centering
  \resizebox{7cm}{!}{\includegraphics{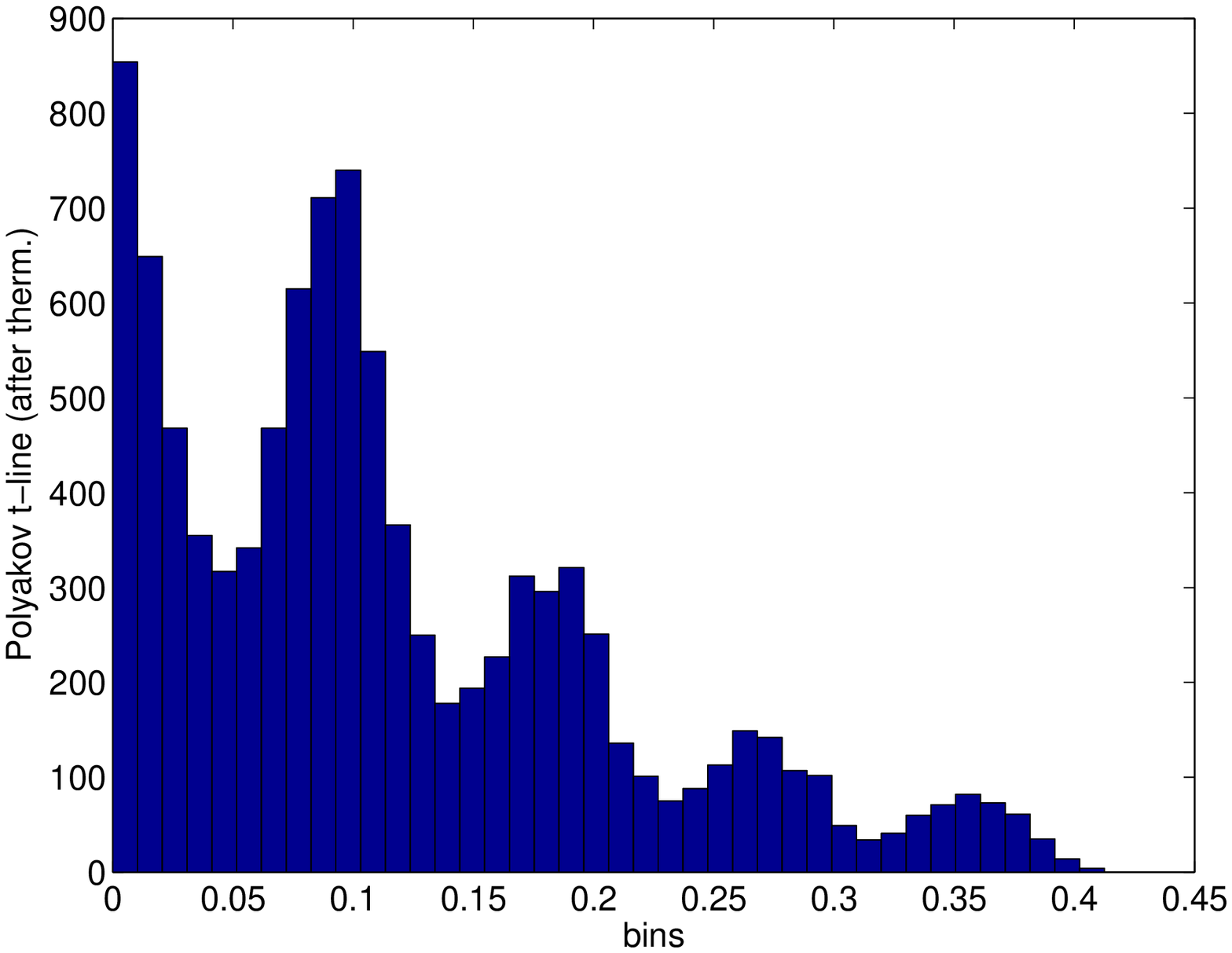}} \ \
  \resizebox{7cm}{!}{\includegraphics{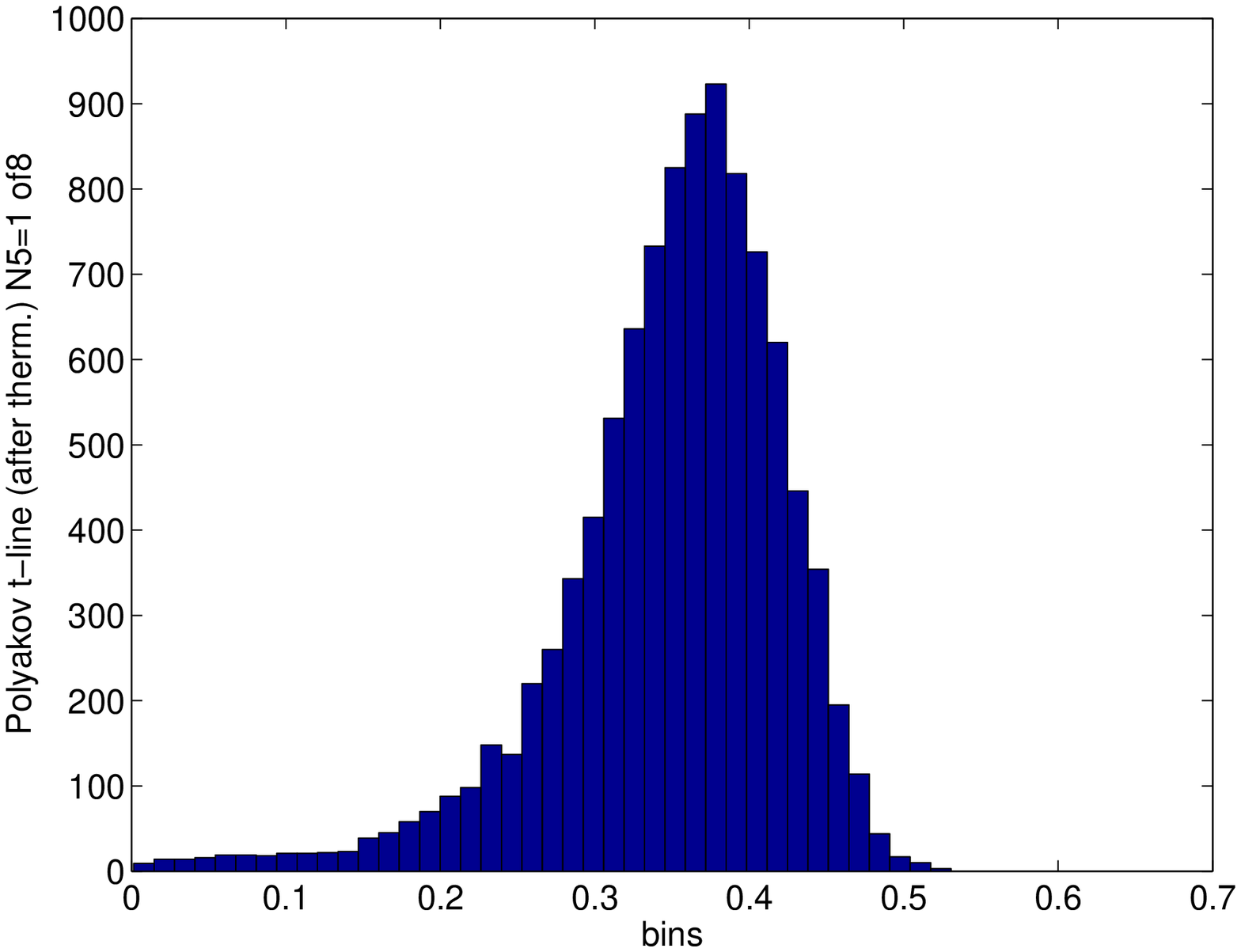}}
  \caption{Histograms of the absolute value of the temporal Polyakov
    loop at $\beta_4=2.32$, $\beta_5=0.5$ and $L_T=4$: taking the average over
    the extra dimension (left plot) or at a fixed slice (right plot)
    \cite{Knechtli:2011gq}.}
  \label{f_histogram}
\end{figure}

At $L_T=4$ we encountered a new phenomenon. If we average the temporal
Polyakov loop over the fifth dimension and then take the absolute value
multiple peaks appear in the histogram of the Monte Carlo history, see the
left plot of \fig{f_histogram}. If instead we take the Polyakov loop at a
fixed slice along the extra dimension, we obtain a single peak, see the
right plot of \fig{f_histogram}. These observations can be interpreted if we
assume that the hyperplanes along the fifth dimension are decoupling from each
other. The finite size scaling analysis using the Polyakov loop at the fixed
slice confirms the universality class of the four-dimensional Ising model also
at $L_T=4$ \cite{Knechtli:2011gq}.

\subsection{Dimensional reduction in the confined phase \label{s_dr}}
\begin{figure}\centering
  \resizebox{7cm}{!}{\includegraphics{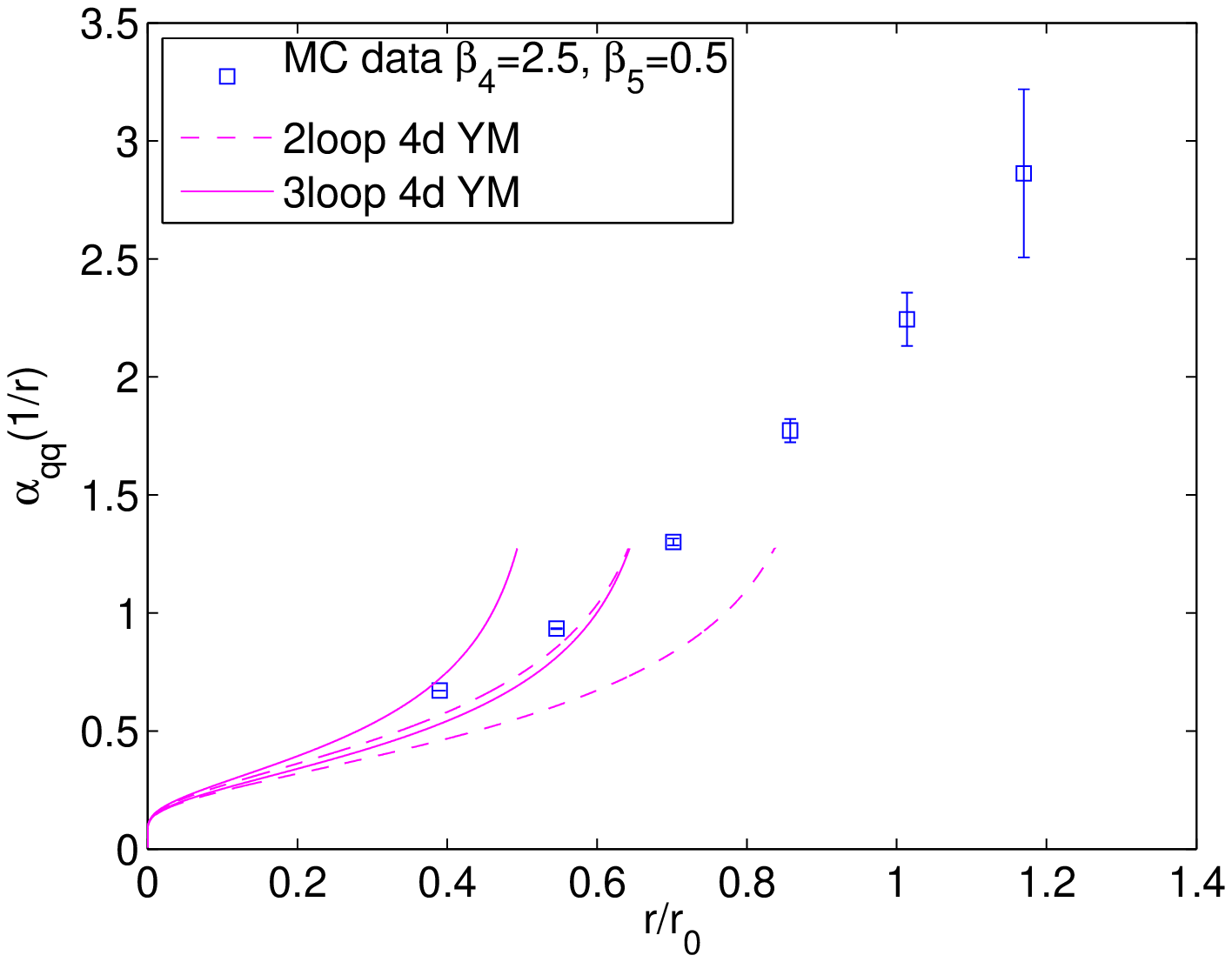}} \ \
  \resizebox{7cm}{!}{\includegraphics{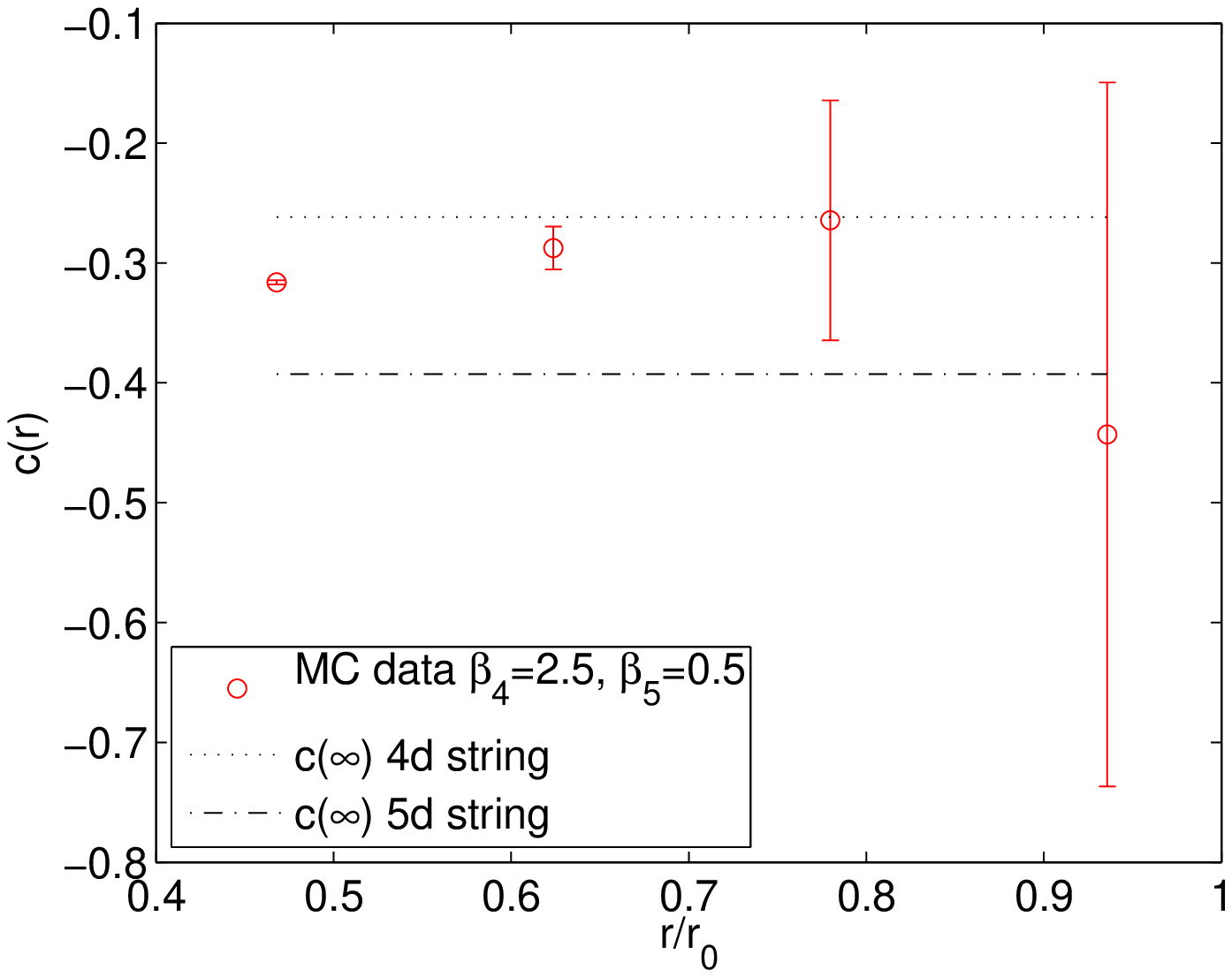}}
  \caption{Shape of the static potential at $\beta_4=2.5$, $\beta_5=0.5$ in
    the confined phase. The coupling $\aqq(1/r)=4r^2F(r)/3$ (left plot) and
    the slope $c(r)=r^3F^\prime(r)/2$ (right plot), where $F$ is the static
    force and the distance $r$ is orthogonal to the fifth dimension.}
  \label{f_couplings}
\end{figure}

We measure the static potential in the five-dimensional confined phase in
large volume. We choose $\beta_4=2.5$ and $\beta_5=0.5$, the lattice
dimensions are $L_T=\Ls=32$, $L_5=16$ and we analyze two replica for a total
of 17234 measurements. Each measurement of the Wilson loops is separated by 10
update iterations and each iteration consists of one heatbath and 16
overrelaxation sweeps through the lattice. For the time-like links we use the
one-link integral \cite{Parisi:1983hm} and for the space-like links 4 levels
of spatial HYP smearing \cite{Hasenfratz:2001hp}. As explained in
\cite{Knechtli:2011gq} we can extract the potential $V(r)$,
where the distance $r$ is taken orthogonal to the fifth dimension. From the
potential we determine the force $F(r)=\{V(r+a_4)-V(r)\}/a_4$ and define the
renormalized couplings
\bea
\aqq(1/r)=4r^2F(r)/3 \quad & \mbox{and} & 
\quad c(r)=r^3F^\prime(r)/2 \,. \label{couplings}
\eea
The effective bosonic string theory \cite{Luscher:1980fr,Luscher:1980ac}
yields the asymptotic value $c(\infty)=-(D-2)\pi/24$, where $D$ is the number
of space-time dimensions.

The lattice spacing measured in units of the scale $r_0$ \cite{Sommer:1993ce}
is $r_0/a_4=6.41(21)$. In \fig{f_couplings} we show our results for the
couplings \eq{couplings}. The data for $\aqq$ are compared to perturbation
theory in the four-dimensional Yang--Mills theory up to three loops and
they are consistent at the two smallest distances. The slope $c(r)$ is harder
to measure, it shows a trend towards the four-dimensional value of the
effective string but the statistical error is already too large at $0.8r_0$. 

\section{Conclusions}

We have explored through Monte Carlo simulations the phase diagram of
five-dimensional SU(2) lattice gauge theory on the torus, in the light of
mean-field results for anisotropy $\gamma<1$.
The phase diagram is shown in \fig{f_phasediag}. The bulk phases which we
identify are the confined and deconfined phase. We do not find a separate
layered phase, like there is in the mean-field phase diagram
\fig{f_meanfield}. Nevertheless we find properties reminiscent of the layered
phase, like the decoupling of the hyperplanes along the fifth dimension and
signs of dimensional reduction in the confined phase. This
could point at a localization mechanism for gauge fields,
cf. \cite{Dvali:1996xe,Laine:2004ji}.

\acknowledgments

This work was funded by the Deutsche Forschungsgemeinschaft (DFG) under
contract KN 947/1-1. In particular A. R. acknowledges full support from the DFG.
The Monte Carlo simulations were carried out on the
Cheops supercomputer at the RRZK computing centre of the University of Cologne
and on the cluster Stromboli at the University of Wuppertal and we thank both
Universities.

\end{document}